# Investigating possible unification of Seyfert galaxies and blazars in *Fermi*-LAT sample

E.U. Iyida[*1,2], I.O. Eya[1,3], F. C. Odo[1,2]

[1]Astronomy and Astrophysics Research Lab., Department of Physics and Astronomy, University of Nigeria, Nsukka
[2]Department of Physics and Astronomy, Faculty of Physical sciences, University of Nigeria, Nsukka
[3]Department of Science Laboratory Technology, Faculty of Physical sciences, University of Nigeria, Nsukka

*email: uzochukwu.iyida@unn.edu.ng

## Abstract

Statistical analyses are invaluable methods used to understand intrinsic emission processes and the unification of extragalactic radio sources. In this paper, we collected radio, X-ray and γ-ray data of blazars from the *Fermi*-LAT and Seyfert galaxies from the INTEGRAL survey and investigated the relationship between the emission properties of Seyfert galaxies and blazar subclasses of flat-spectrum radio quasars (FSRQs) and BL Lac subclasses (BL Lacs). The results show from the average values that these objects follow a sequence that is indicative of probable Seyfert galaxies - BL Lacs - FSRQs unification. We discovered that Seyfert galaxies, BL Lacs and FSRQs share similar emission characteristics in X-ray luminosity ($L_{X\text{-}ray}$), suggestive of fundamental connection while the γ-ray and radio luminosities ($L_{\gamma\text{-}ray}$ and $L_{radio}$) of Seyfert galaxies are the least powerful, signifying an evolving structure. From two-dimensional Kolmogorov-Smirnov test (*K-S* test), we found that Seyfert galaxies differ from the blazar subclasses in $L_{radio}$ while there is no significant difference between them in $L_{X\text{-}ray}$ and $L_{\gamma\text{-}ray}$ which implies that high energy emissions in Seyfert galaxies and blazar subclasses may come from the same emission mechanism. Significant positive correlations exist between the high energy luminosities (X-ray and γ-ray bands) and the low energy component (radio) within the whole sample of blazars and Seyfert galaxies implying a form of connection between them. These results are not only consistent with the prediction of the unified scheme for blazars but also show that Seyfert galaxies have evolutionary link with blazar subclasses.

## 1.0 Introduction

Galaxies are gravitationally bound system of stars, stellar remnant, interstellar gas, dust and even dark matter. They come in different flavours depending on their appearance, spectra, luminosity etc. Galaxies contain supermassive black hole (SMBH) at the center (see, Kormendy & Ho 2013). This SMBH is suckled with enough amount of material within the system making it active with high luminosity across the entire electromagnetic spectrum, thus, entering a phase called the Active Galactic Nuclei (AGNs). These AGNs have highly collimated relativistic jets that are thought to be the product of accretion onto the SMBH (Blandford & Znajek 1977; Blandford & Payne 1982). The jets spread from sub-parsec up to Mega-parsec scales and play essential roles in galaxy evolution (Croton et al. 2006; Heckman & Best 2014). The observational classification of AGNs indicates bimodality in the distribution of their radio-loudness parameter $R_L$ defined as

$$R_L = \log\left(\frac{S_{5GHz}}{S_B}\right) \tag{1}$$

Few of the AGNs (10 - 15%) are radio-loud ($R_L \geq 10$), while the rest (~ 85 %) are radio-quiet ($R_L <$ 10) (Kellermann et al. 1989; Xu et al. 1999; Zhang et al. 2021). Blazars are radio-loud AGNs whose jets are directed close to the line of sight of an observer. They are strong γ-ray emitters (GeV up to TeV) and the radiations from these objects are characterized by apparent superluminal motion in the parsec-scale relativistic jets, large and rapid variability as well as strong polarization (Andruchow et al. 2005; Padovani 2007; Urry 2011; D'Abrusco et al. 2019; Abdollahi et al. 2020).

Depending on their optical properties, blazars are classified into: the BL Lacertae Objects (BL Lacs) with weak emission lines (equivalent width, EW of the emission line in rest frame < 5 Å) and flat-spectrum radio quasars (FSRQs) with strong emission lines, EW ≥ 5 Å) (Stickel et al. 1995; Laurent-Muehleisen et al. 1999; Sbarrato et al. 2014). The ratio of broad emission line luminosity ($L_{BLR}$) to Eddington luminosity ($L_{Edd}$) has also been used to further classify blazars (See, Ghisellini and Celotti, 2001; Ghisellini et al., 2011). FSRQs have $L_{BLR}/L_{Edd} \geq 5 \times 10^{-4}$, while BL Lacs have $L_{BLR}/L_{Edd} < 5 \times 10^{-4}$. This classification implies that blazars differ according to their emission properties. However, based on the synchrotron peak frequencies ($\nu_{s,peak}$ in unit of Hz) of the lower energy bump, BL Lacs are further classified as low (LSPs), with synchrotron peak frequency at far-infrared (far-IR) band, log $\nu_{s,peak}$ < 14, intermediate (ISPs) with synchrotron peak frequency log $\nu_{s,peak}$ between 14 and 15 and high (HSPs) with synchrotron peak frequency log $\nu_{s,peak}$ > 15 (see, Abdo et al. 2010b; Ackermann et al. 2015; Fan et al. 2016).

There is a growing belief that various classes of AGNs may not be intrinsically different but only appear different owing to their orientation relative to the observer's direction. On that note, alternative interpretation of the observational features of these sources has been put forward through the unification scheme. According to the orientation-based unification scheme of jetted-AGNs (see, Urry & Padovani 1995), the parent population of FSRQs are Fanaroff-Riley type II (FR II) radio galaxies. This means that FSRQs are seen when an FR II is observed along its relativistic jet. Similarly, Fanaroff-Riley type I (FR I) radio galaxies constitute the parent population of BL Lacs. This explanation may not be correct, because there are some FR I sources that are connected with FSRQs, and also FR IIs that look like BL Lacs (Urry et al. 1991; Urry & Padovani 1995; Ghisellini et al.

1993). Previous studies seem to offer meaningful proposals that FSRQ samples can metamorphose into BL Lacs via luminosity evolution (Sambruna et al. 1996; Odo et al. 2014; Iyida et al. 2021a). In particular, an evolutionary scenario that relates FSRQs and BL Lacs in terms of the decrease in black hole accretion power with time has been proposed using the analysis from the synchrotron-self and Compton scattered external radiation from blazar jets (Odo et al. 2017). These developments show different trends in the observed properties of FSRQs and BL Lac subclasses across the entire electromagnetic spectrum (Ghisellini et al. 1998; Giommi et al. 2012). Remarkably, it was suggested that the whole blazar subclasses can be arranged in a sequence from HSPs to FSRQs through LSPs and ISPs in order of decreasing synchrotron peak frequencies and increasing source power - the so-called blazar sequence (e.g. Fossati et al. 1998). Thus, in this scenario, the bolometric luminosity of blazars governs the appearance of their spectral energy distributions (SEDs). The most significant of this phenomenon is the negative correlation between the synchrotron peak frequency $v_{s,peak}$ and the synchrotron peak luminosity ($L_{syn}$) of the blazar population. The study of this relationship has been unclear in blazar investigation in a couple of decade and is still under discussion (Fossati et al. 1998; Ghisellini et al. 1998; Ghisellini & Tavecchio 2008; Meyer et al. 2011; Mao et al. 2016; Nalewajko & Gupta 2017; Iyida et al. 2019; Palladino et al. 2019; Pei et al. 2020; Iyida et al. 2021a).

The SEDs of blazars display prominent stability that agrees with the unification scheme, thus, confirming its validity, though the mechanism powering the link among the AGN subclasses is still unclear as their individual emission continuum forms are significantly different across the electromagnetic spectrum (Sambruna et al. 1996; Ghisellini et al. 1998). Even though, the pattern of low and high energy emissions through the leptonic model in ultra-relativistic jets have been used in modeling the SEDs of blazars (see, Iyida et al. 2021b), the role of hadronic processes are yet to be fully understood (see, Iyida et al. 2020). On the other hand, there seems to be a general agreement that appears to support a unification scheme with interesting results indicating orientation links between blazar subclasses and radio galaxies (Fan et al. 2011; Pei et al. 2019, Iyida et al. 2020). Precisely, latest studies using the *Fermi* Large Area Telescope (*Fermi*-LAT) sample suggest that blazar sequence can be extended to radio galaxies by simply invoking an orientation sequence (see, Pei et al. 2020; Iyida et al. 2020; 2021a). All these show that blazar subclasses can be unified as similar objects that differ only in orientation.

Equally, an exceptional type of jetted AGNs that is gaining attention in the unification scheme is the radio-quiet Seyfert galaxy (see, Foschini et al. 2015; Pei et al. 2020; Iyida et al. 2021b). There are growing evidence that these galaxies which host the powerful ultra-relativistic jets with small black hole mass have the tendency to grow to FSRQs or the classical FR IIs (Abdo et al. 2009; 2010a;

Foschini 2017; Paliya et al. 2019). A possible link between Seyfert galaxy and blazar subclasses was first suggested by Oshlack et al. (2001), who remarked how PKS 2004-447 source can be unified with the classical FSRQs. Subsequently, some authors reached the similar conclusion (see, Komossa et al. 2006; Gallo et al. 2006; Yuan et al. 2008; Caccianiga et al., 2014; Schulz et al., 2015). In particular, Gu et al. (2015) carried out Very Large Baseline Array (VLBA) survey on few Seyfert galaxies at 5 GHz, finding out that essentially, Seyfert galaxy with steep radio spectrum has compact morphology that is comparable to radio galaxies. Beside the morphological similarities, there are other hints that point towards the unification of Seyfert galaxies and blazar subclasses. Particularly, a few outstanding Seyfert galaxies are well-known to show blazar properties, for instance, RXJ 16290+4007 (see, Schwope et al. 2000; Grupe et al. 2004) was known as blazar with X-ray and γ-ray emissions (Padovani et al. 2002). This source showed rapid variability in radio, optical and X-ray bands, and was even claimed to be marginally detected in T eV γ-rays (see, Falcone et al. 2004). While a large sample of Seyfert galaxies are known to be classical radio-quiet objects, the radio galaxies are believed to harbour ultra-relativistic jets with extended radio structures that are comparable to Seyfert galaxies (Liu & Zhang 2002; Mathur et al. 2012). Thus, in the revised unification scheme of AGNs, it becomes ideally necessary to embrace these blazar-like galaxies as young jetted counterparts of the classical radio-loud AGNs or rather a portion of larger subset of AGNs that is observed under a specific angle of inclination or geometry. (Singh & Chand 2018). These enigmatic objects present a challenge to current unification scheme of both Seyfert galaxies and blazars and render a unique opportunity to study the emission properties of these AGNs sources. However, an immediate question is whether they are counterparts to radio galaxies that are still under evolution or if they exist as special class of AGNs. What are their general properties in radio, X-ray and γ-ray band? Motivated by these issues, we compiled a sample of *Fermi*-LAT blazars and non-*Fermi*-detected Seyfert galaxies to statistically consider the unification of the classical radio-quiet Seyfert galaxies and blazars. The paper is presented in this way: section 2 discusses the sample selection and analyses. The discussion and the conclusion of this paper are in Sections 3 and 4, respectively.

**2.0 Data Sample and analysis**

From the third *Fermi*-LAT data catalogue compiled by Acero et al. (2015) we selected a sample of 1081 blazars with defined optical identification and luminosities containing 461 FSRQs and 620 BL Lacs. Without considering the unclassified blazars, we collected the monochromatic luminosities of these sources in radio, X-ray and γ-ray frequencies. The clean sample contains 680 blazars (279 FSRQs, 138 HSPs, 133 ISPs and 130 LSPs) with comprehensive information on the three monochromatic luminosities under consideration. However, to enable full study of the emission properties of the current sample in the light of Seyfert galaxy - blazar unification scheme, we included 64 Seyfert galaxies from the INTEGRAL/IBIS survey with measured radio and X-ray luminosities as contained in Chang et al. (2021) and their γ-ray luminosity computed from the information published in Pei et al. (2020b) using $L = 4\pi d_L^2 S(1+z)^{\Gamma-2}$ with $\Gamma$ being the photon spectral index, $z$ and $S$ are the redshift and observed total flux respectively while $d_L$ is the luminosity distance given as $d_L = H_o^{-1} \int \left[(1+z)^2(1+\Omega_m z) - z(2+z)\Omega_\Lambda\right]^{-1/2} dz$. For the current sample, selection effect which is a common signature in AGN activity involving statistical analyses was noted. This was avoided by ensuring that our sample was selected from measurements taken from different instruments and at the same single frequency (monochromatic) of $1.40 \times 10^9$ Hz, $2.52 \times 10^{17}$ Hz and $2.52 \times 10^{23}$ Hz for radio, X-ray and γ-ray respectively. The standard cold dark matter (Λ-CDM) cosmology was adopted throughout the paper with the Hubble's constant $H_0 = 74.20$ kms$^-$Mpc$^{-1}$ and $\Omega_0 = \Omega_m + \Omega_\Lambda$, ($\Omega_{vauum} = 0.70$, $\Omega_{matter} = 0.30$). All relevant data were adjusted based on this concordance cosmology. In the case of statistical analyses, the Pearson Product Moment correlation coefficient ($r$) was used to determine the degree of relationship between these emission parameters using python and MatLab programming softwares.

**2.1 Monochromatic Luminosity Distributions**

We compare using the histogram distributions, the monochromatic luminosities of Seyfert galaxies, FSRQs and BL Lac subclasses in radio, X-ray and γ-ray bands in order to ascertain the level of their relationship in terms of emission mechanisms and as well test the possibility of Seyfert galaxies - BL Lacs - FSRQs unification. The distribution of γ-ray luminosity $L_{γ\text{-ray}}$ (ergs$^{-1}$) and the cumulative distribution function is shown respectively in Figures 1a and 1b. Apparently, the histogram distribution is a continuum with Seyfert galaxies occupying the tail of the distribution while FSRQs are found at the extreme with BL Lacs overlapping Seyfert galaxies and FSRQs with up to 3 orders of magnitude, showing that FSRQs are the strongest γ-ray emitters compared to BL Lacs and Seyfert galaxies. Nonetheless, the distributions yield average (logarithm) values ~ 46.70 ± 0.30 for FSRQ, 45.80 ± 0.20 for LSP, 45.30 ± 0.20 for ISP, 44.30 ± 0.20 for HSP and 43.20 ± 0.30 for Seyfert galaxies. The average values of Seyfert galaxies and blazar subclasses follow the sequence $Log\langle L_{\gamma-ray}\rangle|_{\text{FSRQs}} > Log\langle L_{\gamma-ray}\rangle|_{\text{LSPs}} > Log\langle L_{\gamma-ray}\rangle|_{\text{ISPs}} > Log\langle L_{\gamma-ray}\rangle|_{\text{HSPs}} > Log\langle L_{\gamma-ray}\rangle|_{\text{Seyfert galaxies}}$ which is a signature of Seyfert galaxies - BL Lacs –FSRQs unification**.** Further statistical analysis done on our sample using *Jarque-Bera* test (see, Jarque & Bera, 1980) revealed that FSRQs and Seyfert galaxies are not properly fitted to normal distribution as they skew to the left and right with values of -0.02 and 0.03 respectively. However, in order to examine further, the Seyfert galaxies - BL Lacs -

FSRQs unification using distributive analyses, we applied the Kolmogorov-Smirnov (*K-S*) test. From the cumulative distribution function shown in Figure 1b, it is observed that though, the average sequence exists (indicative of possible unified scheme for these sources) still, there is no significant difference between Seyfert galaxies and the blazar subclasses which implies that γ-ray emission in Seyfert galaxies and blazar subclasses may come from the same emission mechanism. Also, the sequence in the distribution of $L_{γ-ray}$ of these objects is in a sense that supports Seyfert galaxies – blazars unification. The *K-S* test result is shown in Table 1 where *n* is the number of the subsample, $d_{max}$ is the separation distance while *p* is the chance probability.

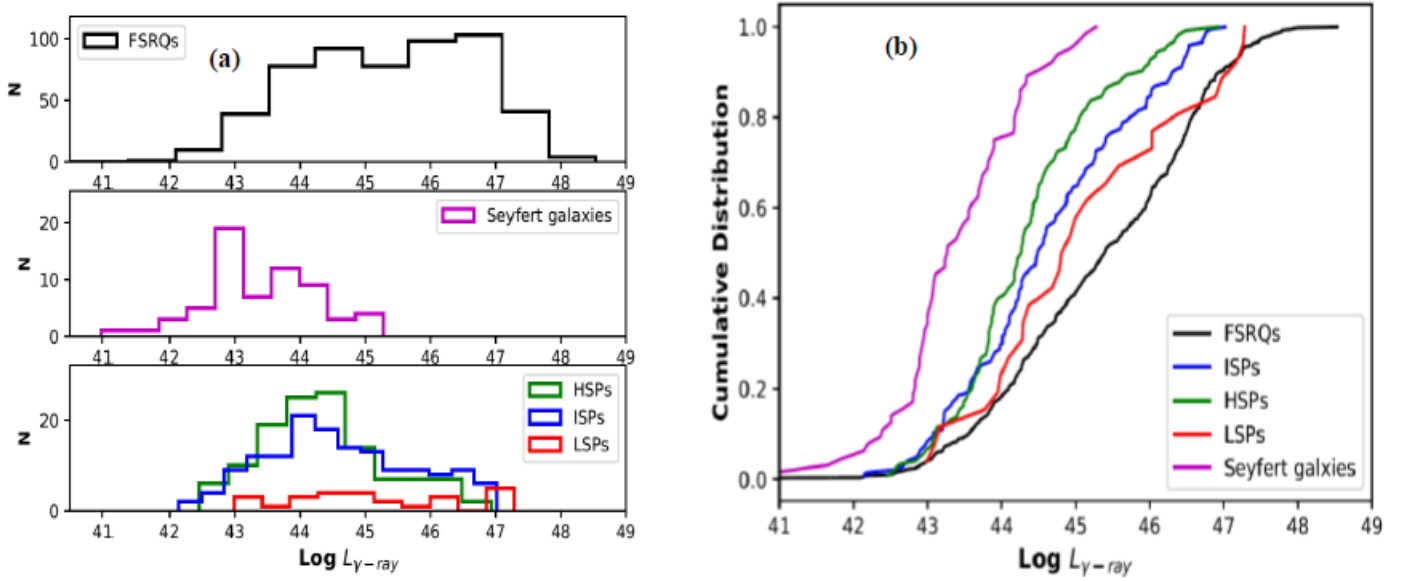

**Figure 1: Histogram showing the comparison of (a) $L_{γ-ray}$ (b) cumulative distribution function of $L_{γ-ray}$ of Seyfert galaxies, FSRQs, and BL Lac subclasses**

The distribution comparing the logarithm of radio luminosity ($L_{radio}$) (in unit of erg/s) of our sample is shown in Figure 2. The $L_{radio}$ of our sample spans several orders of magnitude in an interval of ∼ 36.80 − 45.60 erg s$^{-1}$. FSRQ ranges from 39.20 to 44.80 with a single peak at 43.86 and average value of 42.68 ± 0.20. However, BL Lacs range from 40.20 to 42.60 peaking at different values for ISP, LSP and HSP objects. It is however clear from Fig. 2 (a) that FSRQs, on average, have the tendency to possess higher values of radio luminosity than Seyfert galaxies and BL Lacs. The average value of Seyfert galaxies is 33.80 ± 0.30 while HSPs, ISPs and LSPs have 40.70 ± 0.20, 41.20 ± 0.30 and 42.10 ± 0.10 respectively. Hence, the average values of $L_{radio}$ for our sample follow the relation: $\langle L_{radio} \rangle_{\text{Seyfert galaxies}} < \langle L_{radio} \rangle|_{\text{HSPs}} < \langle L_{radio} \rangle|_{\text{ISPs}} < \langle L_{radio} \rangle|_{\text{LSPs}} < \langle L_{radio} \rangle|_{\text{FSRQs}}$ which is a signature of Seyfert galaxies - BL Lacs – FSRQs unification. A *K-S* test was performed on the data. Results show that generally, there is approximately zero probability that there is a fundamental difference between the distribution of Seyfert galaxies and blazar subclasses in $L_{radio}$. The cumulative distribution function is shown in Fig. 2b while the *K-S* test results are presented in Table 1. This can be interpreted to mean that in terms of $L_{radio}$, different subclasses of BL Lacs may exist and that emissions at different wavebands are important parameters that can be used to investigate Seyfert galaxies – BL Lacs – FSRQs unification scheme.

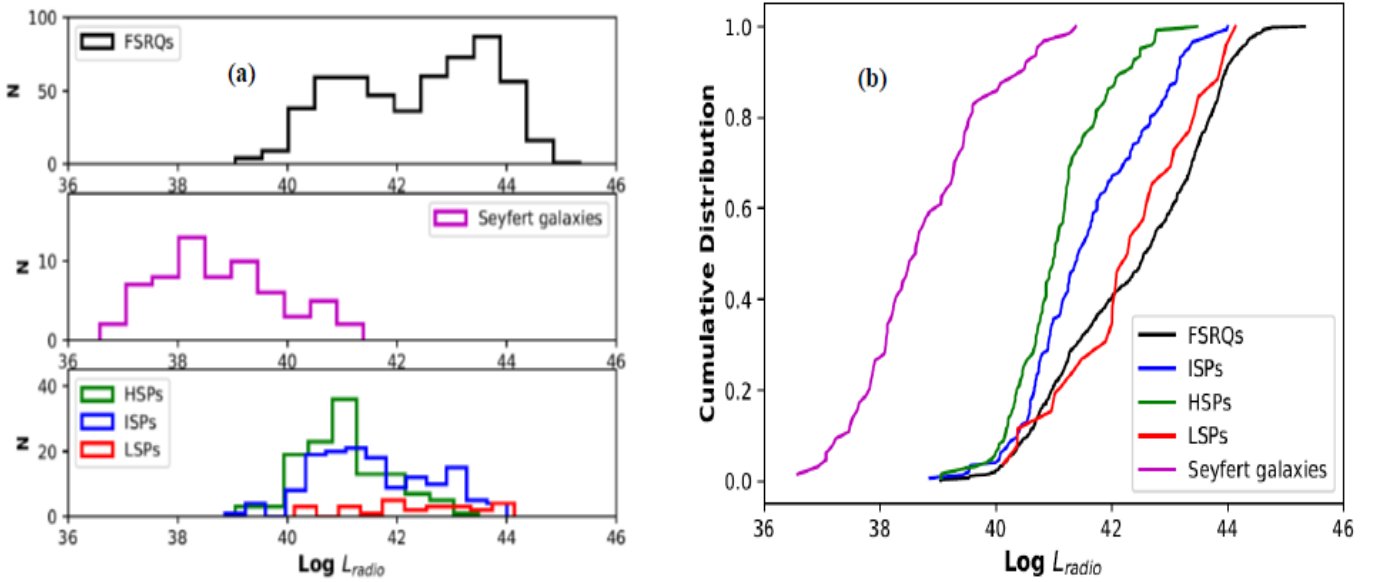

**Figure 2:** Histogram showing the comparison of (a) $L_{radio}$ (b) cumulative distribution function of $L_{radio}$ of Seyfert galaxies, FSRQs, and BL Lac subclasses

To further investigate the unification of Seyfert galaxies and blazar subclasses, we show in Figure 3, the histogram comparing the distributions of the logarithmic of X-ray luminosity ($L_{X-ray}$) for FSRQs, Seyfert galaxies and BL Lac subclasses. While Seyfert galaxies and blazar subclasses occupy similar spaces and generally overlap in $L_{X-ray}$ with up to 3 orders of magnitude implying historic connection, BL Lac objects and FSRQs span the entire range with mean values of 44.60 ± 0.20, 44.21 ± 0.30, 43.20 ± 0.20 and 44.60 ± 0.30 for FSRQs, LSPs, ISPs and HSPs, respectively and 43.50 ± 0.20 for the Seyfert galaxies. However, we noted that for all the distributions, there is continuity with no distinct dichotomy between the subclasses in such a way that is in agreement with the unification scheme, indicating that the subsamples are intrinsically related. A *K–S* test was carried out on the $L_{X-ray}$ data and the results are presented in Table 1. We observed from the cumulative distribution function in Fig.3b that the overlap is very obvious for individual subsamples implying that in general, there is a connection between FSRQs, BL Lacs and Seyfert galaxies.

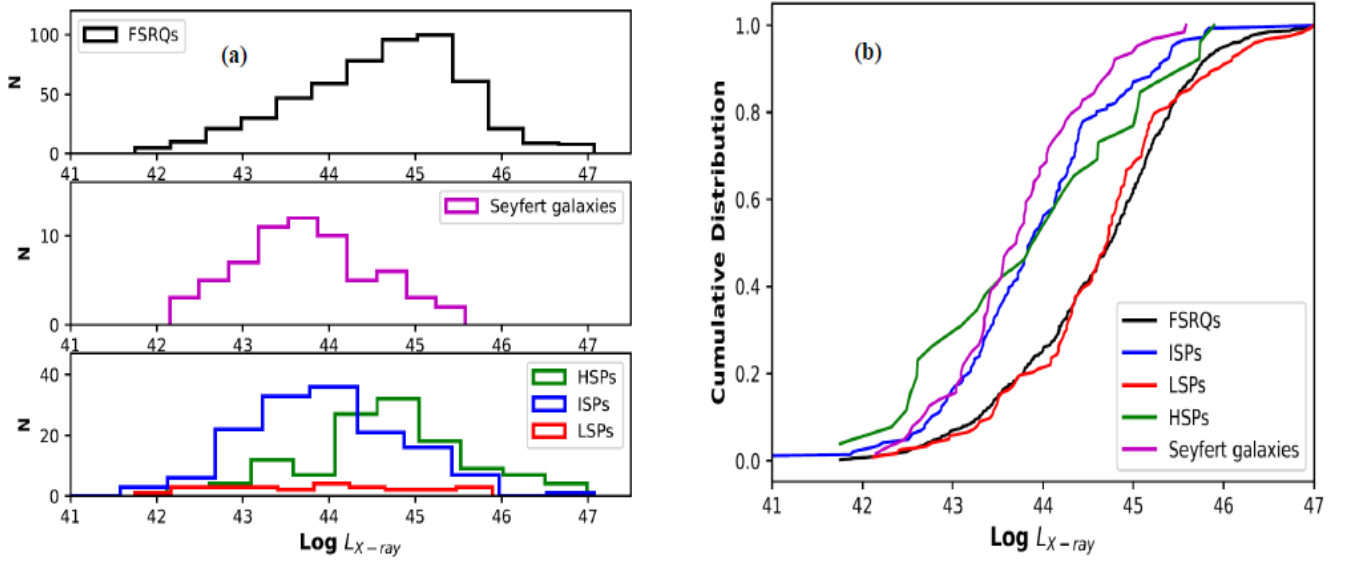

**Figure 3:** Histogram showing the comparison of (a) $L_{X\text{-}ray}$ (b) cumulative distribution function of $L_{X\text{-}ray}$ of Seyfert galaxies, FSRQs, and BL Lac subclasses

**Table 1: Results of *K-S* test of monochromatic luminosities of our sample**

| *Subsamples* | *parameter* | *n* | $d_{max}$ | *p* |
|---|---|---|---|---|
| Seyfert galaxies - HSP | $L_{\gamma\text{-}ray}$ | 64 -138 | 0.54 | $5.20 \times 10^{-06}$ |
| | $L_{radio}$ | 64 -138 | 0.55 | $2.20 \times 10^{-06}$ |
| | $L_{X\text{-}ray}$ | 64 -138 | 0.39 | $2.35 \times 10^{-04}$ |
| Seyfert galaxies - ISP | $L_{\gamma\text{-}ray}$ | 64 -133 | 0.47 | $1.12 \times 10^{-07}$ |
| | $L_{radio}$ | 64 -133 | 0.50 | $3.04 \times 10^{-05}$ |
| | $L_{X\text{-}ray}$ | 64 -133 | 0.22 | $1.32 \times 10^{-05}$ |
| Seyfert galaxies - LSPs | $L_{\gamma\text{-}ray}$ | 64 -130 | 0.33 | $2.08 \times 10^{-05}$ |
| | $L_{radio}$ | 64 -130 | 0.37 | $4.02 \times 10^{-05}$ |
| | $L_{X\text{-}ray}$ | 64 -130 | 0.47 | $3.02 \times 10^{-05}$ |
| Seyfert galaxies -FSRQs | $L_{\gamma\text{-}ray}$ | 64 -279 | 0.67 | $3.32 \times 10^{-06}$ |
| | $L_{radio}$ | 66 -279 | 0.73 | $4.02 \times 10^{-05}$ |
| | $L_{X\text{-}ray}$ | 64 -279 | 0.56 | $2.00 \times 10^{-04}$ |

**2.2 Correlation Analysis**

The existence of correlations between emissions in different wavebands has been used to explain unified scheme of AGNs (see, Pei et al. 2020; Iyida et al. 2021a). The correlation analyses provide valuable information about the emission mechanisms and the connection between the extragalactic sources. We analysed whether the correlations found for Seyfert galaxies, FSRQs and BL Lacs using their parameters are in agreement with the prediction of the general unified scheme. The scatter plot of $L_{\gamma\text{-ray}}$ as a function of $L_{radio}$ is shown in Figure 4. The Seyfert galaxies are not properly aligned with the blazar subclasses in the plot, signifying that they are still in evolution. However, the locations of the subsamples in the plot indicate evolutionary link and possible unification. Table 2 shows the results of linear regression analyses of the subsamples. The slope $a$, intersection $b$, correlation coefficient $r$ and chance probability $p$ and their errors are all listed in the table.

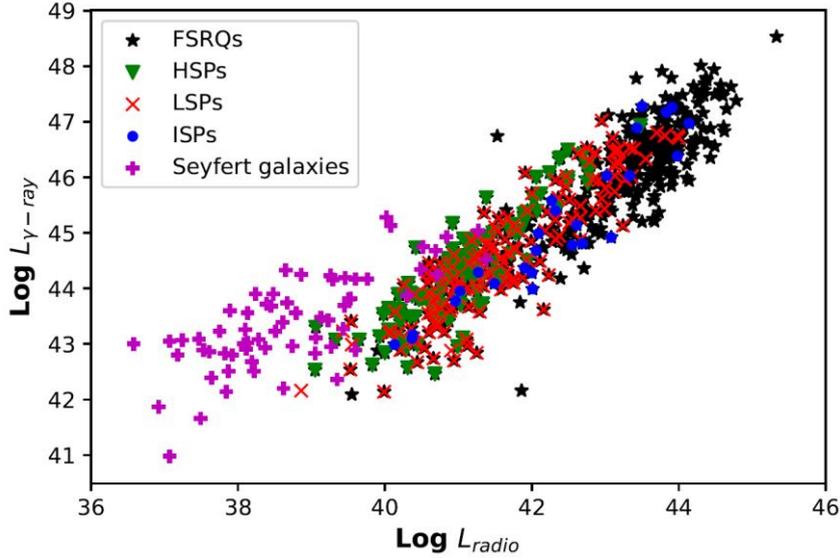

**Figure 4: Plot of $L_{\gamma\text{-ray}}$ against $L_{radio}$ for FSRQs, Seyfert galaxies and BL Lac subclasses**

To consider the form relationship existing between the γ-ray and X-ray emissions of our sample, the $L_{\gamma\text{-ray}}$ - $L_{X\text{-ray}}$ scatter plot is shown in Fig. 5. There is a significant positive correlation for the individual subsamples implying a kind of scaling factor on $L_{\gamma\text{-ray}}$ - $L_{X\text{-ray}}$ plane. We interpret this to mean that similar processes give rise to positive correlation in the different groups at intrinsically different scales. The observed alignment of Seyfert galaxies and blazar subclasses shows that similar processes are responsible for the variations in the properties of these sources which is consistent with the unification scheme. Table 2 shows the results of regression analysis of $L_{\gamma\text{-ray}}$ - $L_{X\text{-ray}}$. The slope $a$, intersection $b$, correlation coefficient $r$ and chance probability $p$ and their errors are all listed in the table. It is noteworthy that the correlations between the subsamples were individually and wholly considered in the analyses.

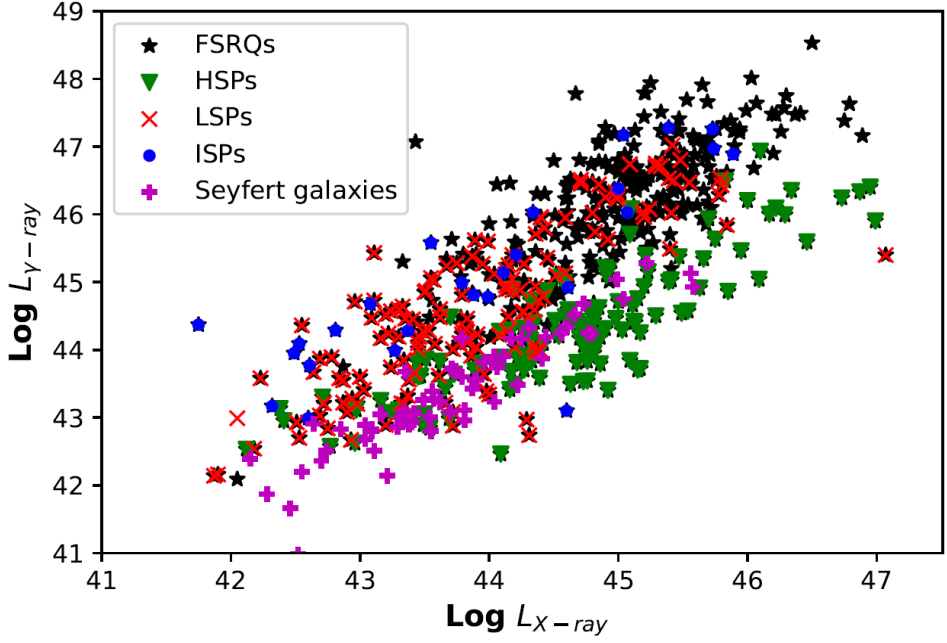

**Figure 5: Plot of $L_{\gamma\text{-}ray}$ against $L_{X\text{-}ray}$ for FSRQs, Seyfert galaxies and BL Lac subclasses**

To investigate the relationships between the emission properties of our sample in X-ray and radio bands, the $L_{X\text{-}ray} - L_{radio}$ plot is shown in Fig. 6. The Seyfert galaxies – BL Lacs – FSRQs sequence is apparent in the plot. Actually, there is a positive correlation of our data for the entire sample of blazars and Seyfert galaxies. This shows that on $L_{X\text{-}ray} - L_{radio}$ plane, these objects follow a trend which is suggestive of a unified scheme. Results of regression analysis of $L_{X\text{-}ray} - L_{radio}$ is shown in table 2. The slope *a*, intersection *b*, correlation coefficient *r* and chance probability *p* and their errors are all listed in the table. It is noteworthy that the correlations between the subsamples were individually and wholly considered in the analyses.

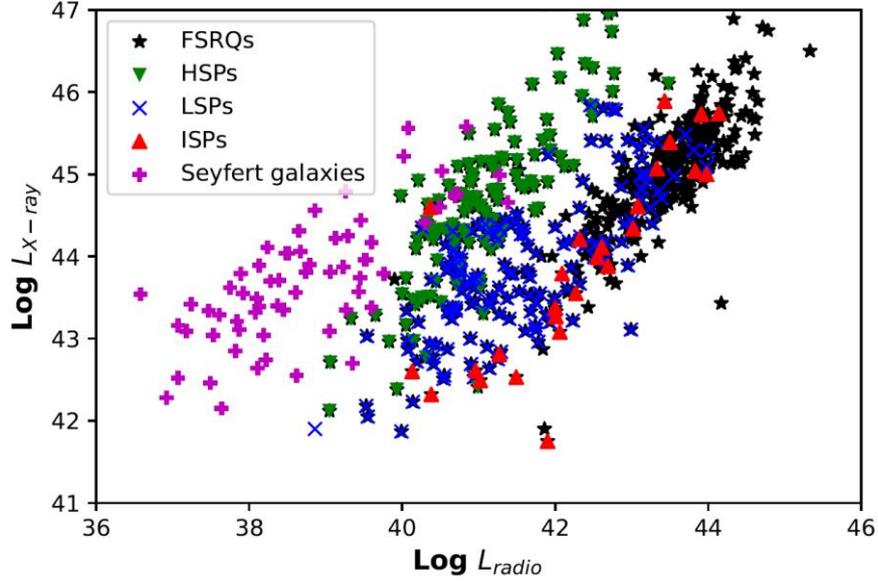

**Figure 6: Plot of $L_{X\text{-}ray}$ against $L_{radio}$ for FSRQs, Seyfert galaxies and BL Lac subclasses**

**Table 2: Results of linear regression analyses of our samples given a** $y = (a \pm \Delta a)x + (b \pm \Delta b)$

| Sample | plots | a | Δa | b | Δb | r | p |
|---|---|---|---|---|---|---|---|
| Whole | $L_{\gamma\text{-}ray} - L_{radio}$ | 0.96 | 0.10 | 42.10 | 0.20 | 0.77 | $2.04\times10^{-09}$ |
| | $L_{\gamma\text{-}ray} - L_{X\text{-}ray}$ | 0.93 | 0.20 | 41.10 | 0.30 | 0.73 | $1.76\times10^{-05}$ |
| | $L_{X\text{-}ray} - L_{radio}$ | 0.82 | 0.20 | 41.21 | 0.30 | 0.55 | $2.47\times10^{-06}$ |
| Seyfert galaxies | $L_{\gamma\text{-}ray} - L_{radio}$ | 0.72 | 0.10 | 42.40 | 0.20 | 0.64 | $5.76\times10^{-07}$ |
| | $L_{\gamma\text{-}ray} - L_{X\text{-}ray}$ | 0.84 | 0.30 | 40.60 | 0.20 | 0.64 | $2.84\times10^{-04}$ |
| | $L_{X\text{-}ray} - L_{radio}$ | 0.72 | 0.10 | 42.40 | 0.20 | 0.64 | $3.71\times10^{-07}$ |
| FSRQs | $L_{\gamma\text{-}ray} - L_{radio}$ | 1.02 | 0.10 | 41.40 | 0.10 | 0.84 | $3.83\times10^{-06}$ |
| | $L_{\gamma\text{-}ray} - L_{X\text{-}ray}$ | 1.02 | 0.20 | 40.60 | 0.30 | 0.74 | $1.58\times10^{-07}$ |
| | $L_{X\text{-}ray} - L_{radio}$ | 0.93 | 0.20 | 40.10 | 0.30 | 0.75 | $3.07\times10^{-06}$ |
| ISPs | $L_{\gamma\text{-}ray} - L_{radio}$ | 1.01 | 0.20 | 41.10 | 0.20 | 0.82 | $2.23\times10^{-08}$ |
| | $L_{\gamma\text{-}ray} - L_{X\text{-}ray}$ | 1.05 | 0.10 | 41.00 | 0.10 | 0.71 | $6.10\times10^{-07}$ |
| | $L_{X\text{-}ray} - L_{radio}$ | 0.90 | 0.30 | 40.40 | 0.10 | 0.71 | $2.32\times10^{-06}$ |
| LSPs | $L_{\gamma\text{-}ray} - L_{radio}$ | 1.03 | 0.10 | 41.30 | 0.30 | 0.81 | $1.14\times10^{-07}$ |
| | $L_{\gamma\text{-}ray} - L_{X\text{-}ray}$ | 1.04 | 0.20 | 41.20 | 0.10 | 0.70 | $4.32\times10^{-07}$ |
| | $L_{X\text{-}ray} - L_{radio}$ | 0.87 | 0.20 | 39.70 | 0.10 | 0.68 | $1.55\times10^{-07}$ |
| HSPs | $L_{\gamma\text{-}ray} - L_{radio}$ | 0.81 | 0.20 | 39.90 | 0.10 | 0.65 | $1.71\times10^{-06}$ |
| | $L_{\gamma\text{-}ray} - L_{X\text{-}ray}$ | 0.59 | 0.10 | 40.20 | 0.20 | 0.68 | $5.01\times10^{-05}$ |
| | $L_{X\text{-}ray} - L_{radio}$ | 0.63 | 0.10 | 41.30 | 0.20 | 0.62 | $1.10\times10^{-04}$ |

## 2.3 Effect of redshift on monochromatic luminosities of our sample

The luminosity calculation suggests that there is an apparent mutual band luminosity correlation since the luminosity of AGNs subclasses are known to be redshift-dependent. To investigate for the real correlations among the luminosities in different wavebands of our sample, the effect of redshift (z) has to be completely removed. This was done by using the Spearmann's partial correlation equation (see, Padovani et al. 1992; Odo and Ubachukwu, 2013) given as

$$r_{ab,c} = \frac{r_{ab} - r_{ac}r_{bc}}{\sqrt{(1-r_{ac}^2)(1-r_{bc}^2)}},  \quad (2)$$

where $r_{ab}$, $r_{ac}$ and $r_{bc}$ represent the correlation coefficients between two variables (luminosities) $x_a$ and $x_b$, $x_a$ and $x_c$ or $x_b$ and $x_c$ respectively while $r_{ab,c}$ is the partial correlation coefficient between two variables with $z$ dependence removed. When equation (2) was applied to our data, the Spearmann's test results are shown in Table 3. the column (1) gives the plot, column (2) gives the class of the sample, column (3) is the correlation coefficient, column (4) is the correlation coefficient between two variables $x_a$ and $x_b$, $x_a$ and $x_c$ or $x_b$ and $x_c$ with the redshift effect, column (5) is correlation coefficient between the variables $x_a$ and $x_b$, $x_a$ and $x_c$ or $x_b$ and $x_c$ with the redshift $z$, column (6) is the correlation coefficient with the redshift effect removed, column (7) gives the chance probability for the two distribution to come from the same distribution. Therefore, after the redshift effect is completely removed, there is still correlation between among the luminosities for the Seyfert galaxies and blazar subclasses. Thus, we argue that the correlations are not driven by redshift effects, rather they are intrinsically induced.

Table 3: Results of *Regression analysis after removing the redshift effect*

| Sample | plots | r | $r_{ab}$ | $r_{bc}$ | $r_{abc}$ | p |
|---|---|---|---|---|---|---|
| Whole | $L_{\gamma\text{-ray}} - L_{radio}$ | 0.77 | 0.63 | 0.59 | 0.73 | $3.03\times10^{-05}$ |
| | $L_{\gamma\text{-ray}} - L_{X\text{-ray}}$ | 0.73 | 0.64 | 0.63 | 0.78 | $3.10\times10^{-07}$ |
| | $L_{X\text{-ray}} - L_{radio}$ | 0.55 | 0.52 | 0.60 | 0.58 | $2.30\times10^{-06}$ |
| Seyfert galaxies | $L_{\gamma\text{-ray}} - L_{radio}$ | 0.69 | 0.67 | 0.63 | 0.65 | $2.12\times10^{-04}$ |
| | $L_{\gamma\text{-ray}} - L_{X\text{-ray}}$ | 0.64 | 0.59 | 0.54 | 0.66 | $1.21\times10^{-06}$ |
| | $L_{X\text{-ray}} - L_{radio}$ | 0.64 | 0.53 | 0.58 | 0.69 | $1.02\times10^{-09}$ |
| FSRQs | $L_{\gamma\text{-ray}} - L_{radio}$ | 0.84 | 0.74 | 0.68 | 0.80 | $4.08\times10^{-05}$ |
| | $L_{\gamma\text{-ray}} - L_{X\text{-ray}}$ | 0.74 | 0.67 | 0.63 | 0.68 | $4.43\times10^{-06}$ |
| | $L_{X\text{-ray}} - L_{radio}$ | 0.75 | 0.66 | 0.58 | 0.74 | $3.20\times10^{-06}$ |
| ISPs | $L_{\gamma\text{-ray}} - L_{radio}$ | 0.82 | 0.76 | 0.72 | 0.81 | $1.23\times10^{-03}$ |
| | $L_{\gamma\text{-ray}} - L_{X\text{-ray}}$ | 0.71 | 0.69 | 0.64 | 0.72 | $1.65\times10^{-04}$ |
| | $L_{X\text{-ray}} - L_{radio}$ | 0.71 | 0.72 | 0.69 | 0.73 | $2.93\times10^{-04}$ |
| LSPs | $L_{\gamma\text{-ray}} - L_{radio}$ | 0.81 | 0.73 | 0.50 | 0.84 | $2.20\times10^{-05}$ |
| | $L_{\gamma\text{-ray}} - L_{X\text{-ray}}$ | 0.70 | 0.72 | 0.63 | 0.73 | $2.90\times10^{-04}$ |
| | $L_{X\text{-ray}} - L_{radio}$ | 0.68 | 0.56 | 0.62 | 0.65 | $8.09\times10^{-05}$ |
| HSPs | $L_{\gamma\text{-ray}} - L_{radio}$ | 0.65 | 0.59 | 0.67 | 0.64 | $2.05\times10^{-06}$ |
| | $L_{\gamma\text{-ray}} - L_{X\text{-ray}}$ | 0.68 | 0.61 | 0.56 | 0.57 | $1.03\times10^{-05}$ |
| | $L_{X\text{-ray}} - L_{radio}$ | 0.62 | 0.62 | 0.57 | 0.60 | $4.09\times10^{-04}$ |

## 3. Discussion

The unification scheme known for its simplicity has become popular because of the promise it holds in bringing two apparent quite distant extragalactic objects under one roof. We have investigated the possible unification of blazars and Seyfert galaxies in terms of their emission properties in different wavebands. The existence of this relation between the emission properties of blazars and Seyfert galaxies has opened new scenarios on the comprehension of the unification scheme for these extragalactic radio sources. (see, Gallo et al. 2003; Falcke et al. 2004; Iyida et al. 2021b). The unification scheme proposes that FSRQs and BL Lacs are different expressions of the same physical process that vary only by bolometric luminosity (Ghisellini et al. 1998: Fossati et al. 1998). Consequently, there should be continuity in distributions of the $L_{radio}$, $L_{X-ray}$ and $L_{\gamma-ray}$ of these objects.

Remarkably, the comparison of the distributions of $L_{\gamma-ray}$ and $L_{X-ray}$ of the present sample does evidently show that FSRQs are the extreme versions of Seyfert galaxies. The average values of Seyfert galaxies and blazar subclasses follow a sequence that is indicative of unification scheme. The Seyfert galaxies are somewhat occupying the lowest regime of $L_{\gamma-ray}$ and $L_{radio}$ while FSRQs occupy the highest regime of the distribution with BL Lacs being intermediate in the configuration. This is actually in agreement with evolutionary connection in which AGNs may start off as Seyfert galaxy in the low energy regime and grow with different luminosity passing through BL Lacs to the extreme FSRQs, suggesting that Seyfert galaxies are the young jetted AGNs in evolution (e.g. Singh & Chand 2018). Thus, it can be argued from the distributions of $L_{radio}$ and $L_{\gamma-ray}$ that emission mechanism may not be the major difference between Seyfert galaxies and blazar subclasses. In addition, if this argument is correct, then one can expect that the distributions of the $L_{X-ray}$ for blazars and galaxies should be from the same parent distribution. The *K-S* test results of $L_{\gamma-ray}$ and $L_{X-ray}$ given in Table 1 implies that Seyfert galaxies and blazar subclasses may come from the same emission mechanism as there is no significant difference among them, implying that they can be unified.

Also, the new unified scheme of AGNs posits that these blazar-like galaxies called Seyfert galaxies are young jetted counterparts of traditional radio-loud AGNs or a part of a larger AGN subclass observed under a specific geometry and inclination of the line of sight (Singh & Chand, 2018). Thus, can be unified along with the classical blazar subclasses. In this sense, there should be continuity in distributions of the properties of Seyfert galaxies, BL Lacs and FSRQs. The distributions of the $L_{radio}$ and $L_{X-ray}$ of current sample of blazars and Seyfert galaxies are apparently in agreement with the scheme, as there are no clear dividing lines between the subclasses of blazars and Seyfert galaxies which signifies that these sources are the same but differ only in orientation. The interpretation is that the main mechanism that is generating the emissions in our sample are the same, though, they differ systematically among the AGNs subclasses (Chen et al. 2016), thus, indicating an evolutionary link among these AGNs subclasses which is in agreement with the unification scheme as earlier suggested by Fossati et al. (1998) and Ghisellini et al. (1998).

Moreover, while there is strong correlation between $L_{\gamma-ray}$ and both $L_{radio}$ and $L_{X-ray}$ for Seyfert galaxie, BL Lacs and FSRQs, thus, implying that these sources can be unified. Chen et al. (2016) found significant correlation between $L_{\gamma-ray}$ and $L_{radio}$ of blazars showing that γ-ray emission scales linearly with radio and X-ray properties of these sources. The difference between Seyfert galaxies, FSRQs and BL Lacs in this study, can be due to orientation as there is continuity in the distributions of these sources. We interpret this to mean that the mechanisms producing emissions in γ-ray and X-ray of our subsamples are not different (e.g. Chen et al. 2016). For the Seyfert galaxies in the case of the $L_{radio}$- $L_{X-ray}$ plane, we argue that any γ-ray emission is linked to the fact that they are still in evolution. Using a sample of X-ray data of BL Lacs, Odo et al. (2012) obtained a similar result and argued that X-ray emission is directly linked to radio emission for the BL Lacs in their sample, which this our case, we included many other AGNs in our current sample. Perhaps, the strong correlations obtained for the $L_{γ-ray}$ –$L_{radio}$ and $L_{γ-ray}$ –$L_{X-ray}$ data are indication that these sources can be unified, in which case, Seyfert galaxies differs from FSRQs and BL Lacs by orientation. In addition, after completely removing the common dependence of luminosity on redshift, the $L_{radio}$ and $L_{X-ray}$ still show strong positive correlation with partial Spearman correlation coefficient ($r > 0.60$) at 99.9 % confidence level. The implication of the significant correlations is that Seyfert galaxies, FSRQs and BL Lacs have the emission condition though differ by orientation angle, thus, can be unified.

**4.0 Conclusion**

We have selected a complete sample of blazars from the *Fermi*-LAT database and Seyfert galaxies from the INTEGRAL survey in order to study the relationship between these sources in the radio, X-ray and γ-ray bands and consider if they can be unified. We observed from histogram distributions of $L_{radio}$ and $L_{γ-ray}$ that there is sequence with Seyfert galaxies having least emissions while FSRQs have the strongest while BL Lacs have intermediate values. The sequence of emission processes is consistent with the prediction of the unification scheme. The distributions of the observed parameters of our sample suggest a smooth transition from Seyfert galaxies to FSRQs through LSPs, ISPs and HSPs. Our results thus strongly support the hypothesis of Seyfert galaxies –BL Lacs –FSRQs unification. We found the existence of significant positive correlations ($r > 0.60$) between Lγ-ray - Lradio and Lγ-ray – LX-ray within the whole sample and individual subsamples. The significant correlations imply that these extragalactic sources are intrinsically the same class of objects though differ by emitting power, thus, can be unified.

**Acknowledgement**

We sincerely thank an anonymous referee whose invaluable comments and suggestions helped to improve the manuscript.